%% file: main.tex
\documentclass[sigconf]{acmart}
\usepackage{color}
\usepackage{xcolor}
\usepackage{tikz}
\usepackage{pgfplots}
\usepgfplotslibrary{groupplots}
\usepackage{graphicx}
\usepackage{tabularx}
\usepackage{listings}
\usepackage{amsmath}
\usepackage{graphicx}

\AtBeginDocument{%
  }

\pgfplotsset{compat=1.18}

\begin{document}

\title{Building a P2P RDF Store for Edge Devices}

\author{Xuanchi Guo}
\email{xuanchi.gou@tu-berlin.de}
\orcid{1234-5678-9012}
\affiliation{
  \institution{Technical University of Berlin}
  \country{Germany}
}

\author{Anh Le-Tuan}
\email{anh.letuan@tu-berlin.de}
\orcid{0000-0003-2458-607X}
\affiliation{
  \institution{Technical University of Berlin}
  \country{Germany}
}

\author{Danh Le-Phuoc}
\email{danh.lephuoc@tu-berlin.de}
\orcid{0000-0003-2480-9261}
\affiliation{
  \institution{Technical University of Berlin}
  \country{Germany}
}

\renewcommand{\shortauthors}{AAA et al.}

\input{00Abstract}



\keywords{The Semantic Web, Peer-To-Peer system, Distributed RDF Store, Edge Devices}


\maketitle

\input{01Introduction}
\input{05RelatedWorks}

\input{02Design}

\input{03Architecture}
\input{04Evaluation}

\input{06Conclusion}
\begin{acks}
This work is supported by the German Research Foundation (DFG) under the COSMO project (grant No. 453130567), and by the European Union's Horizon WINDERA under the grant agreement No. 101079214 (AIoTwin), and RIA research and innovation programme under the grant agreement No. 101092908 (SmartEdge). 
\end{acks}

\bibliographystyle{ACM-Reference-Format}
\bibliography{bibliography}

%\appendix

\end{document}

%% file: 00Abstract.tex
% The structured abstract should contain the following four sections:

% Purpose: Explain ‘why’ you undertook this study. If presenting new or novel research, explain the problem that you have solved. If building upon previous research, briefly explain why you felt it was important to do so.
% Methodology: Explain ‘how’ you did it. Let readers know exactly what you have done to reach the results. For example, did you undertake interviews? Did you experiment in the lab? What tools, methods, protocols, or datasets did you use?
% Findings: Explain ‘what’ you found during the study, whether it answers the problem you set out to explore, and whether your hypothesis was confirmed. You need to be very clear and direct and give exact figures rather than generalise.
% Value: Provide readers with an analysis of the value of your results. You can also conjecture what future research steps could be.

\begin{abstract}
The Semantic Web technologies have been used in the Internet of Things (IoT) to facilitate data interoperability and address data heterogeneity issues. The Resource Description Framework (RDF) model is employed in the integration of IoT data, with RDF engines serving as gateways for semantic integration. However, storing and querying RDF data obtained from distributed sources across a dynamic network of edge devices presents a challenging task. The distributed nature of the edge shares similarities with Peer-to-Peer (P2P) systems. These similarities include attributes like node heterogeneity, limited availability, and resources. The nodes primarily undertake tasks related to data storage and processing. Therefore, the P2P models appear to present an attractive approach for constructing distributed RDF stores. Based on P-Grid, a data indexing mechanism for load balancing and range query processing in P2P systems, this paper proposes a design for storing and sharing RDF data on P2P networks of low-cost edge devices. Our design aims to integrate both P-Grid and an edge-based RDF storage solution, RDF4Led for building an P2P RDF engine. This integration can maintain RDF data access and query processing while scaling with increasing data and network size. We demonstrated the scaling behavior of our implementation on a P2P network, involving up to 16 nodes of Raspberry Pi 4 devices.
\end{abstract}

%% file: 01Introduction.tex
\section{Introduction}
The emergence of the Internet of Things (IoT) has enabled communications between physical and virtual devices, as they can connect to the network without direct human intervention. Communication and data exchanges occur between many IoT devices within an IoT deployment. Nonetheless, a significant challenge that hinders real-life IoT deployment is the lack of data interoperability~\cite{ali2015internet}. Data interoperability is the ability of various components within an IoT deployment to share and understand data. To achieve data interoperability, IoT systems need data integration and meaningful understanding capabilities to handle different data formats and semantics. 

The Semantic Web technologies, which aim to provide interoperability for data on the Web, offer several solutions to address data heterogeneity issues in the IoT domain~\cite{chatzimichail2021semantic}. For instance, heterogeneous IoT data is integrated using the RDF data model, which standardises how metadata descriptions and the underlying data of Web-based resources are defined and used~\cite{shi2018survey}~\cite{gravina2018integration}. In addition, RDF engines act as semantic integration gateways for IoT data~\cite{kiljander2014semantic}. 

To tackle the network latency problem between cloud and end-user devices in cloud-based centralised real-time IoT deployments, decentralised IoT edge architectures have been put forward~\cite{9032541}. These architectures shift data processing to the edge of the IoT, close to edge devices and sensors. Thus, edge devices can collect and compute IoT data instead of sending them back to a central site or cloud. Moreover, these architectures offer several advantages to IoT platforms and devices, such as reducing communication latency and unnecessary network bandwidth and improving operational efficiency.

As the decentralised integration paradigm fits better with the distributed nature of the autonomous deployment of smart IoT devices, RDF4Led~\cite{le2018rdf4led} was proposed to move RDF data processing to IoT edge devices. The RDF engine consists of an RDF storage and SPARQL processor tailored for a lightweight edge machine to store and query RDF data. RDF4Led could store up to 100 million RDF statements on a common edge device by minimising memory consumption and maximising data capability. However, like other trends in edge computing that outsource RDF engines close to the edge of the network, RDF4Led does not consider the computational resources of adjacent edge devices. To process rapidly growing RDF data efficiently at scale, edge-based RDF engines must adopt a distributed infrastructure that adheres to the Semantic Web's inherently distributed nature while avoiding centralised RDF repositories' drawbacks.

\begin{figure*}[ht]
    \centering
    \includegraphics[width=0.8\textwidth]{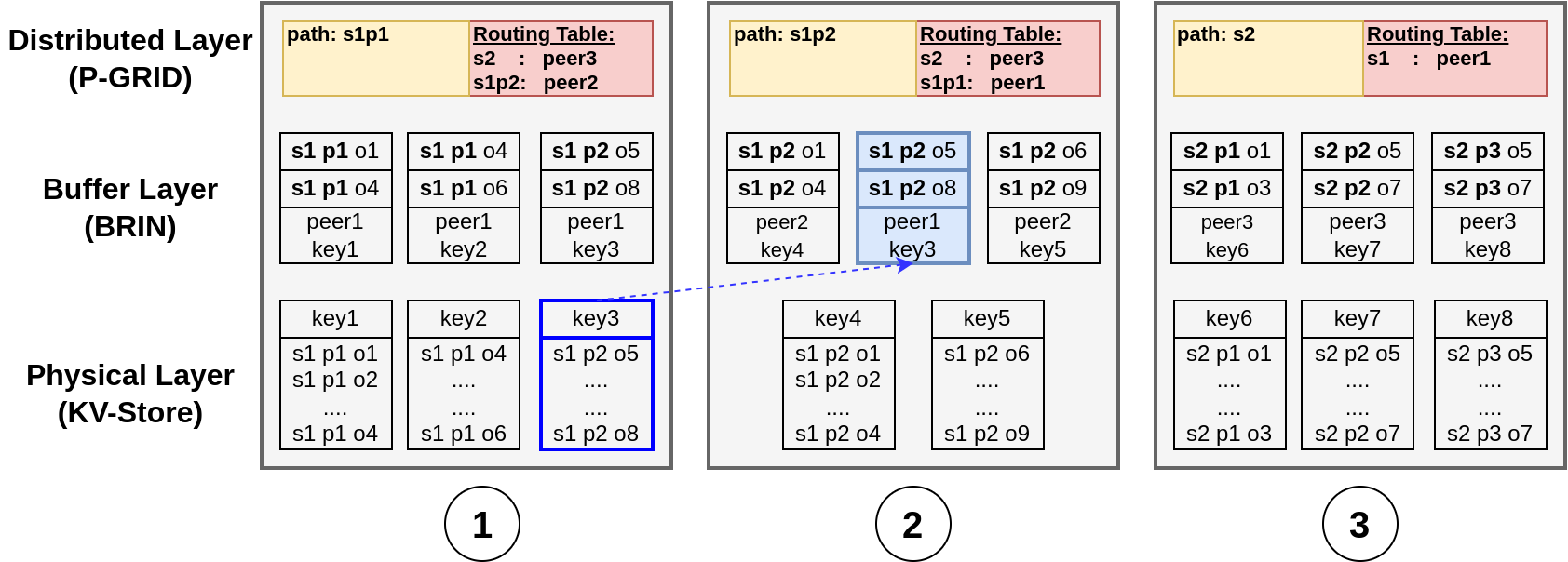}
    \caption{An example of the three-layer organisation of SPO layout. Each circled number represents a peer in the network. The dotted blue line represents that the data block has a replica on the buffer layer of another peer.}
    \label{fig:three-layer}
\end{figure*}

Processing highly distributed RDF data over adjacent edge devices would pose challenges, such as locating relevant data sources and balancing workload among nodes. 
One of the reasons for the challenges is the inflexibility of the client-server communication model when deployed at the edge of the IoT~\cite{Marx:2018}. 
The Peer-to-Peer (P2P) communication model has been argued as a suitable solution to manage distributed data. P2P is a well-known communication model in which each node (or peer) acts as a server and a client because it can send requests and responses to other nodes (peers)~\cite{milojicic2002peer}. P2P has also shown great potential in building highly distributed platforms for decentralised applications and data management for ever-growing increasing information on the Web~\cite{milojicic2002peer}. 
Furthermore, the P2P model would fit well with IoT edge scenarios, which often contain distributed edge devices~\cite{karagiannis2019edge}. It can support the implementation of distributed edge-based applications by equipping edge devices with the capability to cooperate to achieve common goals. The P2P model for edge computing can leverage many edge devices' computational and storage resources. It can also offer flexibility for dynamic edge networks and enhance information sharing between edge nodes~\cite{karagiannis2019edge}. 

This motivates us to use a P2P model to build an RDF store for lightweight edge devices to manage and process large-scale RDF data efficiently. P-Grid~\cite{aberer2001gridella} is a structured P2P system that provides load balancing and efficient search using randomised routing. Besides, it abstracts a trie structure, which makes it suitable for processing range queries commonly used in RDF data querying. Notwithstanding, its original design does not support the RDF data model and edge devices. Meanwhile, RDF4Led was developed as a lightweight RDF storage and SPARQL processor that is tailor-built for edge hardware. 
Consequently, integrating P-Grid and RDF4Led can provide a promising solution to create a decentralised architecture to store and share RDF data on the edge of the IoT. On top of the RDF4Led storage design, we add an additional index layer to enable indexing on the P2P system.

The contributions of this paper are as follows:
\begin{enumerate}

    \item An alternative design for a distributed RDF store for the P2P system of edge devices based on RDF4Led and P-Grid. 

    \item A complete implementation of a distributed RDF store in Java by integrating the P-Grid and RDF4Led code base.

    \item A set of experiments to evaluate the performance of the implementation in a P2P system using numerous Raspberry Pi 4 devices. Measurement and analysis of the time taken to search and join operations of the RDF data under different data sizes and network sizes. 
    
\end{enumerate}
This paper is constructed as follows. Section~\ref{section2} explains the rationale of the system from the aspects of storage design and access structure. Section~\ref{section3} describes the architecture of the system and its detailed implementation. The experimental evaluation of the system is discussed in Section~\ref{section4}. Section~\ref{section5} discusses the related work and the paper is summarised in Section~\ref{seciton6} with conclusions and future work.

%% file: 05RelatedWorks.tex
\section{Related Work}
\label{section5}
Federated query processing approaches are widely used for querying distributed RDF data across multiple heterogeneous data sources. These approaches decompose each query into subqueries directed to the SPARQL endpoints of related data sources and retrieve the results in an integrated manner~\cite{oguz2015federated}. Despite providing complete query results, query federation introduces a single point of failure and faces challenges in efficiently managing a large number of data sources and queries due to the execution of subqueries on multiple data sources.

To address scalability issues, some research works have explored the combination of RDF data storage with a P2P architecture, with a focus on RDF data indexing and query processing within these networks. Peers collaborate to build a distributed index and achieve optimal load balancing for storage and query tasks.

In the realm of decentralised architectures for sharing and querying semantic data, Piqnic~\cite{10.1007/978-3-030-21348-0_1} stands out as a resilient and decentralised solution. By employing replication, Piqnic ensures data availability and resilience to node failures. However, it falls into the category of unstructured P2P systems, where queries are flooded throughout the network, leading to challenges such as low search efficiency, lack of guarantee for rare data retrieval, and increased network traffic.

In contrast, structured P2P systems offer several advantages, including scalability, robustness, load balancing, and predictable searching costs for distributed RDF data stores. Research efforts like RDFPeers~\cite{cai2004rdfpeers}, 3rdf~\cite{ali20113rdf}, and Atlas~\cite{koubarakis2006semantic} use DHT-based P2P overlays for distributed RDF data storage and querying. 
RDFPeers~\cite{cai2004rdfpeers} is the pioneering P2P system that implements a distributed RDF repository. It stores each triple at three different places in the network and can handle various native queries, including atomic triple patterns, disjunctive and range queries, and conjunctive multi-predicate queries. However, RDFPeers has inherent limitations, including challenges in load balancing mechanisms for peers storing popular triples and the lack of support for data indexing strategies tailored to edge devices.

Inspired by the advantages of structured P2P RDF repositories, our work leverages the state-of-the-art structured P2P system, P-Grid, to extend RDF4Led for large-scale, lightweight device networks. The integration aims to address the challenges associated with querying RDF data in decentralised environments, providing promising opportunities for scalable and efficient query processing in edge applications.

 %RDF4Led is a scalable single-node RDF engine for lightweight edge devices that operates with low memory consumption and fast updates. However, it does not support indexing RDF data across adjunct peers in a network. Meanwhile, P-Grid is a self-organising structured P2P system that can handle arbitrary key distributions and provide storage load-balancing and efficient search. 

%% file: 02Design.tex
\section{Distributed RDF Storage Using P-Grid Model}
\label{section2}

\subsection{Design of Distributed RDF Storage}
To design distributed RDF engines for the P2P system of lightweight edge devices, we adopt the RISC-style design philosophy in~\cite{neumann2008rdf}. The features of an RDF store are centralised around data access and join operations. To answer a SPARQL query, the primary mission of an RDF engine is to perform graph pattern matching over the RDF dataset. The RDF engine has to search for RDF triples that match triple query patterns and compute the joins between the matched triples. In the scope of this paper, we aim to enable enhanced data access for RDF data in a P2P environment and reuse the join operators in state-of-the-art engines such as RDF4Led. That means we focus on indexing RDF data in a P2P system of edge devices to find triples that match a triple pattern efficiently. 

%We leave further improvement of the join operators for future work.
%In this section, we will 
%This section covers the design of the three-layer distributed RDF data store, which essentially adds a distributed layer to the existing two-layer storage of RDF4Led. Each layer of the system will be discussed through a brief illustration. 

RDF data can be stored with multiple indexes; thus, different triple query pattern variants can be efficiently answered~\cite{harth2005optimized}. 
The multiple indexes approach ensures that whichever components of a triple pattern are bound, there is always an appropriate index for an efficient search for the triples that match the pattern. 
Hence, we organise RDF triples in three indexing layouts: \emph{SPO} (Subject - Predicate - Object), \emph{POS}, and \emph{OSP}.
These three permutations are sufficient to answer all query patterns, e.g., the SPO layout can cover the triple patterns with the bound subject $(s, ?p, ?o)$ and the bound subject-predicate $(s, p, ?o)$.  
%Creating all six access paths as in Hexastore~\cite{weiss2008hexastore} may answer complex queries more effectively. However, considering the storage consumption for lightweight edge devices and the maintenance update cost in a P2P network, three index layouts are more suitable than six.

We use a hybrid three-layer indexing strategy to maintain the index for RDF triples in a P2P system, including Physical Layer, Buffer Layer, and Distributed Layer. According to the RDF4Led storage design, the Physical Layer involves storing RDF data in flash storage. The Buffer Layer is used to cache recently accessed data and data updates before reading from or writing to the Physical Storage and index data in the Physical Storage. The Distributed Layer defines how the RDF data is distributed over the decentralised P2P network utilising a P-Grid overlay structure. Figure~\ref{fig:three-layer} illustrates an example of an SPO layout composed of these three layers in our system.

The Physical Layer can be viewed as a key-value store. RDF graphs are compressed into numerous RDF molecules, which are compact sorted lists of properties and objects related to one subject as described in~\cite{le2018rdf4led}.
Therefore, storage space could be greatly saved by avoiding redundant storage of subject values. These RDF molecules are sorted into pages and then grouped into blocks, which adapt to the flash I/O behaviour. In Figure~\ref{fig:three-layer}, it is assumed that RDF triples are stored as encoded binary strings, and subscripts of the triples represent the order of binary strings for simplicity. To illustrate, $(s_1, p_1, o_1)$ is before $(s_1, p_1, o_2)$ as $o_1$'s encoded string is smaller than $o_2$'s encoded string. Peer1 stores three key-value pairs\emph{(molecules)} in its physical layer. The molecule with \emph{key1} uses its first tuple as its self key, and its value is the combination of the ordered tuples from $(s_1, p_1, o_1)$ to $(s_1, p_1, o_4)$.

Regarding the index structure in the Buffer Layer, RDF4Led adopts a similar idea to Block Range Index(BRIN) using a small tuple to represent the information of data blocks from its persistent storage. This approach minimises the memory size to store and maintains the index data. In the middle of Figure~\ref{fig:three-layer}, each peer has a Buffer Layer with data blocks related to key-value pairs in its Physical Layer. Peer1's first data block is formed by extracting the first tuple $(s_1, p_1, o_1)$ and the last tuple$(s_1, p_1, o_4)$ as well as the key \emph{key1} of its first key-value pair, that is the first RDF molecule stored in its Physical Layer. Each data block also indicates its original owner. For example, the blue data block of Peer2 points to an RDF molecule stored in Peer1's Physical Layer. Because Peer1 is the actual owner who fully holds the molecule, Peer2 merely has this RDF molecule's summarised information. 

The Distributed Layer is a virtual overlay layer running on top of the physical network. It is formed by building a fully decentralised access structure P-Grid based on the Distributed Hash Table(DHT) abstraction. Like the other DHT-based P2P systems, P-Grid links each RDF peer with partitions of the overall RDF data space. Thus, it enables the decentralised storage and maintenance of RDF data among RDF peers. The distribution of RDF data and RDF peers in the P-Grid overlay is exemplified in the top layer of Figure~\ref{fig:three-layer}. It shows a particular case where the overall RDF graph is partitioned into exactly three parts, each of which denotes RDF triples starting with $s_1p_1$, $s_1p_2$, or $s_2$. Each peer has a path associated with what data partition it owns in its storage.

Additionally, 
% it can be noticed that the tree structure in this example is a full binary tree, in which each node except the leaf node has either none or two offspring, and its leaf levels are not at the same depth. Besides, 
Peer1's path is a concatenation of binary strings of $s_1$ and $p_1$ referred to as $s_1p_1$. All RDF triples (except triples stored as replicas) in its storage start with $s_1p_1$.

Moreover, each peer owns a routing table to which it could look up whom to forward queries when the requested data is out of its range. Thanks to its routing table, Peer1 is aware that when it is queried for RDF triples starting with $s_2$, it should forward these queries to Peer3. Furthermore, for queries regarding $s_1p_2$, Peer2 may have the requested data. Thus, Peer1 will send the query to Peer2 instead. 

\subsection{P-Grid Access Structure}
\label{subsection:access structure}
In this subsection, we will focus on the access structure of the P-Grid model with the illustration of a specific example. Moreover, we will also introduce P-Grid's prefix-based routing scheme in detail. 

Though ~\cite{punceva2003efficient} states that access structures using k-ary balanced trees can significantly reduce the number of hops compared to binary trees, we assume the binary tree structure is constructed. This assumption conforms with the fact that RDF triples are encoded in binary strings. 

In accordance with~\cite{10.1007/3-540-44751-2_15}~\cite{datta2005range}~\cite{aberer2002scalable}, each RDF peer has a unique address that identifies itself in the community of peers. It can use this address to communicate with other peers in the network. That means there is a one-to-one mapping between peer \textit{x} and its address \textit{addr}: $x \mapsto addr$, where $addr$ belongs to the full address space $ADDR$.

Different from the original key space of the P-Grid access structure, we define the maximal key length as a fixed number $m$. If each element of a triple is encoded into an integer 32-bit long, the maximal key length $m$ of an RDF triple is 96. A binary string $key_t$ represents the key of an RDF triple $t$
\begin{equation*}
    key_t=p_1p_2...p_k, 1 \leq k \leq m,
\end{equation*}
where each digit $p_i \in \{0,1\}$ .
The value of each key is the sum of all non-zero exponents of 2 :  
\begin{equation*}
    val(key_t)=\sum_{i=1}^{k}p_i2^{m-i}.
\end{equation*}
Additionally, interval 
\begin{equation*}
    I(key)=[val(key), val(key)+2^{m-k}) \subseteq K,
\end{equation*}
Each interval $I(key)$ indicates a key space partition and $K$ denotes the entire key space.

One of the distinguishing features of the P-Grid is that its peers' identifiers are decoupled from their paths. They do not have constant or predetermined paths in the overlay network. Their paths vary dynamically during the network construction and maintenance for more balanced data distribution. The data partition determines each peer's path it is responsible. It also tells the peer's location in the overlay network. Assume that peer \emph{p} stores a set of data items $\delta(p)$ and each data item is encoded into a key, then $\delta(p)$ is a set of all these keys,
\begin{equation*}
    \delta(p) = \{key_1, key_2, ... , key_p\}.
\end{equation*}
The path of peer $p$ is defined as the common prefix of all keys,  
\begin{equation*}
    path(p) = p_1p_2...p_n.
\end{equation*}
Each path is mapped to a key interval. In other words, we could learn from $path(p) = p_1p_2...p_n$ that peer $p$ takes responsibility for the interval $I(path(p))$ in the key space and all keys starting with $p_1p_2...p_n$ fall under peer $p$'s key space. 
Note that although P-Grid has the abstraction of a tree, the nodes residing in the overlay network are hierarchy-less and are all leaf nodes in the tree. 

\begin{figure}[h]
    \centering
    \includegraphics[width=\linewidth]{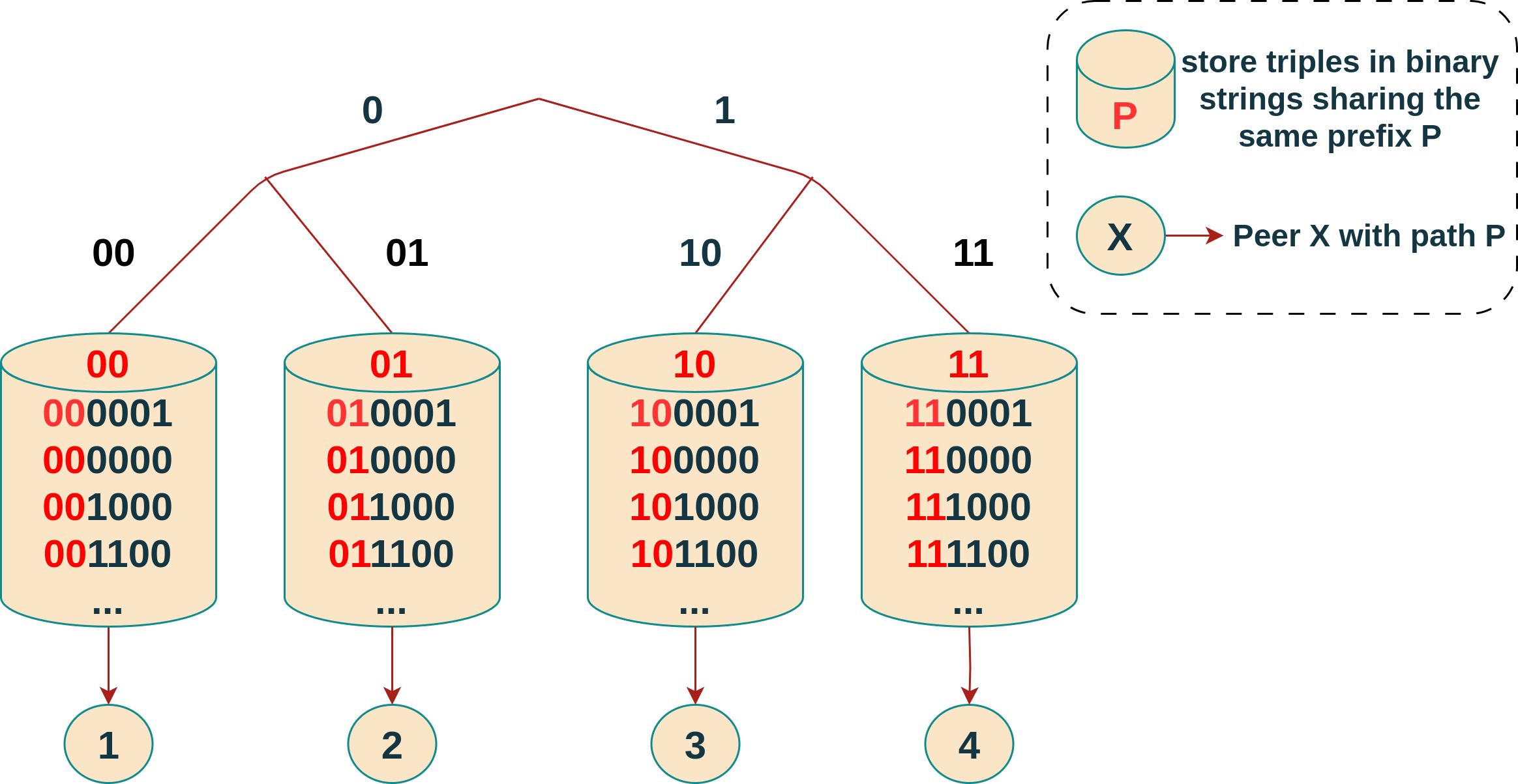}
    \caption{Example of P-Grid Trie Structure, showing four peers in a perfect binary search tree with a maximum level of two. The binary strings represent the encoded RDF triples stored on each peer.}
    \label{fig:P-Grid Trie Structure}
\end{figure}

Take Figure~\ref{fig:P-Grid Trie Structure} as an example; the trie has four peers and binary strings represent all RDF triples. Peer1 stores RDF triples like: 
$(s_1, p_1, o_1)$,
$(s_1, p_1, o_2)$,
$(s_1, p_2, o_3)$,
$(s_1, p_4, o_6)$,
After being encoded, these triples are transformed into fixed-length binary strings(here, we assume the length is 6 for simplification), which in turn will be:
   \textbf{00}0001,
    \textbf{00}0000,
    \textbf{00}1000,
    \textbf{00}1100.
As these keys share the common prefix \textbf{00}, Peer1's path is defined as \textbf{00}. The same holds for other peers in the network.

Because of its trie structure, P-Grid's searching algorithm is based on a prefix routing scheme. Each peer maintains a routing table. Each level of the routing table contains one or multiple references to a peer on the other side of the binary tree at the same level. The entry level denotes the prefix length. For its prefix with length $i$:
\begin{equation*}
   prefix(p,i)=p_1p_2...p_i, 1 \leq i \leq n-1
\end{equation*}
, and $prefix(p,0)$ is empty, peer $p$ keeps references to other peers in its routing table:
\begin{equation*}
   ref(p,i)= I_i=\{p_x \parallel \forall p_x, prefix(p_x, i) = prefix(p, i-1)+p_i^\neg \}. 
\end{equation*}
Thus, peer $p$ keeps a list of $n$ entries $(0, I_0),..., (n-1, I_{n-1})$ as its routing table. The peers in $I_i$ have the same prefix of length $i-1$, but its digit at position $i$ is opposite to that of $path(p)$.

\begin{figure}[h]
    \centering
    \includegraphics[width=\linewidth]{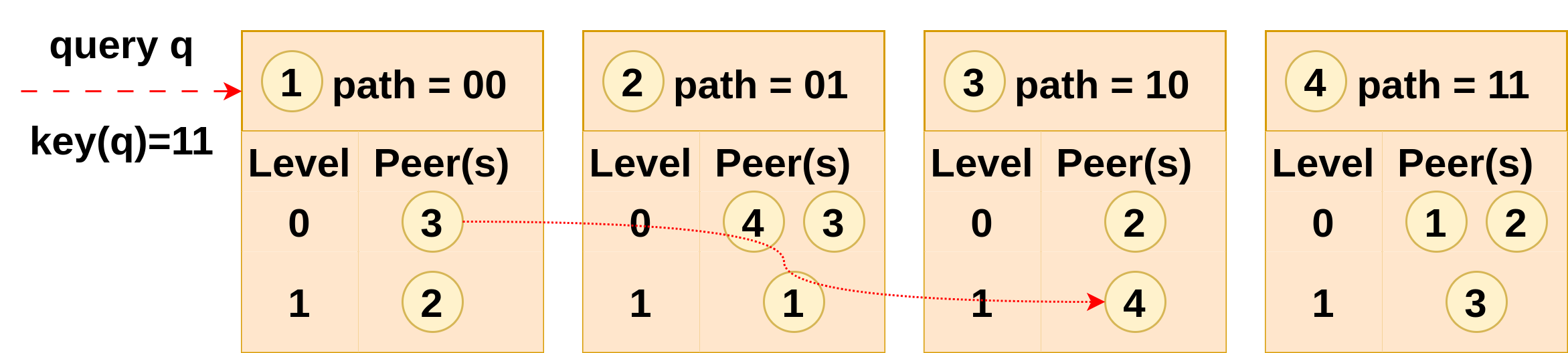}
    \caption{Example of P-Grid Routing Table, corresponding to the P-Grid trie structure in Figure~\ref{fig:P-Grid Trie Structure}. Each peer's routing table has a maximum level of 2. The dotted directed red lines show the paths that a query, whose path is 11, follows to find the answering peer using the prefix-based routing mechanism. }
    \label{fig:P-Grid Routing Table}
\end{figure}

Since all peers have paths of length 2 in the binary tree shown in Figure~\ref{fig:P-Grid Trie Structure}, their routing tables' highest level is 2 (indexing from 0) as shown in Figure~\ref{fig:P-Grid Routing Table}. Take the routing table of Peer1 for an example; at level 0, it stores Peer3 or Peer4 or both as the reference peers; at level 1, Peer2 is selected accordingly.

The routing table ensures that a peer will answer a query as long as its requesting data exists in the overlay. Peer1 can only answer queries with key \textbf{00}. When it is required to answer a query $q$ with key \textbf{11}, Peer1 learns from its routing table quickly that it should forward this query to Peer3. Because $q$ and Peer1's path have an empty common prefix, Peer3 is the only candidate at level 0. By being forwarded to the next peer, $q$ is getting closer to its final destination. After that, Peer3 forwards the query $q$ to Peer4. Because $q$ and Peer3's path have a common prefix of length 1, Peer4 is the candidate at level 1. Finally, Peer4 receives the query, which is within its scope. It will forward its locally searching results back to the query initiator.

%% file: 03Architecture.tex
\section{System Architecture and Implementation}
\label{section3}
\begin{figure}[ht!]
    \centering
    \includegraphics[width=\linewidth]{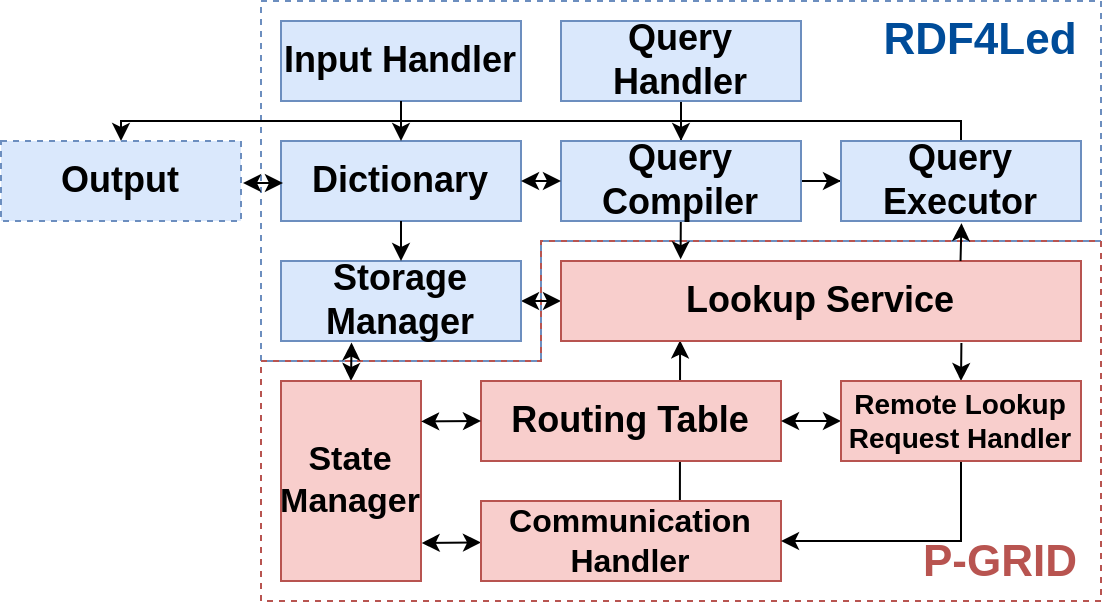}
    \caption{Overview of system architecture, showing the relationship between the major components. Each component is taken from RDF4Led or P-Grid. Each arrow pointed in a direction indicates a dependency relationship between the modules. }
    \label{fig:architecture}
\end{figure}

This section will %mainly 
describe our implementation that integrates the RDF4Led engine and P-Grid system to create a distributed RDF store for a P2P network of lightweight edge devices. 
It utilises the flash-friendly RDF storage of RDF4Led and the P-Grid virtual binary search tree to efficiently manage and query RDF data on each peer in the network.
Figure~\ref{fig:architecture} illustrates the architecture %overview that 
integrating the RDF4Led and P-Grid components on a single peer.
The critical components %integration of the design is 
to be extended are the RDF storage and SPARQL query processor of RDF4Led, and the State Management and Lookup Service of P-Grid.

Here, the blue part represents the original architecture of RDF4Led consisting of an \emph{Input Handler} that is tied to a \emph{Dictionary} to translate between string-based RDF resources and encoded identifiers. 
\emph{Dictionary} adopts a hash function to create a fixed-length integer deterministically as a representation of an original string of arbitrary length. 
Because of its natural behaviour, the hash function is suitable for key-value structures. 

The encoded RDF triples are indexed with three index layouts (SPO, POS, OSP) and are stored with a \emph{Storage Manager} that employs a two-layer index for each layout as presented in~\cite{le2018rdf4led}.
SPARQL queries are registered on the system via a \emph{Query Handler} and are compiled with a \emph{Query Compiler}. For compiling a SPARQL query, the \emph{Dictionary} will involve converting RDF nodes in basic graph patterns to encoded identifiers. A \emph{Query Executor} is implemented to execute the query plans computed by the \emph{Query Compiler} and to produce the output results. The \emph{Output Handler} returns the original format of RDF resources for these output results from the \emph{Query Executor}.

%--------------------------------------------------
%State manager
The red part encompasses essential functions adopted from P-Grid~\cite{aberer2003p}. The \emph{State Manager} from P-Grid serves as a controller for a peer, facilitating state transitions based on given inputs. It includes primary states, such as the bootstrapping phase, exchange phase, replicating phase, and running phase.
The bootstrapping phase initiates when a peer joins the P2P network, aiming to discover and familiarise itself with other participants. Subsequently, during the exchange phase, existing peers in the P-Grid overlay structure undergo stabilisation, but data distribution might remain imbalanced. To address this, the exchange phase reorganises and sorts data items among RDF peers.
A static approach with a global replication factor of two ensures that each data item has two replicas in the P2P network. During the exchange phase, only data blocks are replicated, with each replica recording the origin peer containing the actual RDF triples within the block. Origin peers halt initiating replicating requests until their data blocks meet the global replication requirement.
Once the exchange phase is complete, the running phase commences, making a peer ready to work. Peers in this phase can both initiate query requests and respond to queries from other peers in the P-Grid network.
Throughout each phase, the \emph{State Manager} communicates through a \emph{Communication Handler}, facilitating message exchange. The \emph{Lookup Service} triggers lookup requests to the \emph{Remote Lookup Request Handler}, which forwards requests to other peers. The \emph{Routing Table} aids the \emph{State Manager} and the \emph{Remote Lookup Request Handler} in identifying the peers to communicate with.

%--------------------------------------------------
%storage manager
With this architecture, each peer in the network has an RDF4Led \emph{Storage Manager} responsible for storing and maintaining the RDF data locally. The \emph{Storage Manager} handles data insertion or deletion and resolves query requests. If new data needs insertion or updating, the \emph{Dictionary} will first encode the string into an identifier to accelerate the search and save the memory space in the \emph{Storage Manager}. The design of the flash-aware storage layout and indexing scheme of a single RDF4Led machine are in use as they cater to the need for a suitable storage method for lightweight edge devices. Hence, the \emph{Storage Manager} contains a buffer layer and a physical RDF storage layer. The data in the physical layer is organised as data blocks; the buffer layer is the index of each data block in the physical layer. In our system, the indexes of the data blocks are published to the \emph{State Manager}. Using the peer information from the \emph{Routing Table} and based on the indexed key, the \emph{State Manager} will decide which data block should be replicated or exchanged to which peer to maintain the load balance for the network. 
To retrieve RDF triples from the Physical Layer, the \emph{Storage Manager} initially searches the Buffer Layer to identify the indexes of the data blocks potentially containing the desired results. Subsequently, the \emph{Storage Manager} accesses the encoded values from the Physical Storage Layer, utilising the key value of each data block. This retrieval allows the \emph{Storage Manager} to further decompose the encoded value into multiple tuples, facilitating subsequent result trimming.

%--------------------------------------------------
%lookup service
After compiling a SPARQL query, the \emph{Query Compiler} computes an optimal query plan. Each triple query request of the query plan is resolved by the \emph{Lookup Service}, which will search in the local storage of a peer or forward the request to remote peers. The search mechanism in the P2P system is indicated by \emph{Routing Table}, which is essential for a structured P2P overlay, as it holds the information of other peers. The \emph{Routing Table} ensures that a triple query request is answered by a particular peer as long as the requested data exists in the overlay. When matched triples are found in a peer, the result sets are forwarded back to the \emph{Query Executor} as a final or intermediate result. The final result generated by the \emph{Query Executor} would be translated by the \emph{Dictionary} back to the original format of the triples as the output.

%% file: 04Evaluation.tex
\section{Evaluation and Analysis}
\label{section4}
\input{Chapter04/41Setup}

\input{Chapter04/42Experiments}

%% file: Chapter04/41Setup.tex
\subsection{Evaluation Setup}

\subsubsection{Software and Hardware}

%\hfill\\

%The distributed RDF engine was implemented in Java, and we reused parts of the source code from RDF4Led and P-Grid, including the RDF dictionary from RDF4Led and the bootstrap module from P-Grid. 
We implemented our system in Java and reused as much of the source code from RDF4Led and P-Grid as possible. We also re-implemented some parts using updated technologies. For instance, we recycled the dictionary module from RDF4Led and the bootstrapping mechanism from P-Grid.
%P-Grid connectivity was initially implemented with Java WebSocket, which could not handle asynchronous message passing. In the newer version, we implemented the connection module with gRPC~\footnote{https://grpc.io/}. 
The Java WebSocket implementation in the initial version of P-Grid was replaced by gRPC to improve the system's ability to handle asynchronous message passing.

%Our experiments were conducted on a cluster of 4 to 16 Raspberry Pi 4 (Pi4) devices, representing lightweight and low-cost edge devices for the IoT. Each device is equipped with quad-core processors at 1.5GHz, 8GB RAM, and an onboard LAN speed of 1Gbps.
We conducted our experiments using a cluster of 4 to 16 Raspberry Pi 4 (Pi4) devices, which serve as lightweight and cost-effective edge devices for the IoT. Each device is equipped with quad-core processors clocked at 1.5GHz, 8GB of RAM, and an onboard LAN connection with a speed of 1Gbps. Peers are considered directly interconnected with every other peer in the experiments.

%\subsubsection{Assumptions}
%Through performing simulation on the algorithm of constructing P-Grid,
%has explored 
%several questions have been explored regarding exchanging algorithms and achieved a simple and intuitive argument from their simulation, that the exchange function inherently tends to balance the distribution of keys
%~\cite{10.1007/3-540-44751-2_15} .

%By performing simulations on the algorithm for constructing P-Grid, several questions have been explored regarding exchanging algorithms. The simulations have yielded a simple and intuitive argument, indicating that the exchange function inherently tends to balance the distribution of keys ~\cite{10.1007/3-540-44751-2_15}.
%Therefore, we assume that in the following experiments, the P-Grid construction process has halted, resulting in a reasonably balanced distribution of keys. 

\subsubsection{Performance Metrics}
%\hfill\\
In this evaluation, we focus on testing and evaluating our system's performance in terms of query execution time (QET). The metric is critical in edge applications where data access and retrieval within numerous lightweight computing devices are of paramount importance.
%In evaluating the performance of our RDF engine, we focus on testing and evaluating the RDF engine's performance in terms of query execution time (QET). This metric is crucial in edge applications, where fast data access and retrieval with numerous lightweight computing devices are a top priority.
Throughout our evaluation process, we measured the QET of searching and retrieving matching RDF triples of an atomic triple pattern among a set of P2P nodes, as well as the QET of join operations across multiple atomic query patterns.

\subsubsection{Dataset and Storage setup}
%\hfill\\
For our experiments, we utilise the ISD (Integrated Surface Dataset)~\footnote{\url{https://www.ncdc.noaa.gov/isd}}, a notable weather dataset comprising weather observations collected from 20 thousand weather stations worldwide since 1901. This dataset encompasses various measurements, including temperature, wind speed, wind angle, and more. Moreover, each observation is accompanied by timestamps indicating when these measurements were recorded. 
%Figure~\ref{fig:NOAA schema} is adapted from~\cite{le2020pushing} with permission from the authors, depicting the RDF schema of a sensor reading.
%We use the SSN/SOSA ontology~\cite{10.3233/SW-180320} to describe the sensor metadata and sensor readings of the ISD dataset, the data schema which were utilized in our previous work~\cite{le2020pushing}.

To transform the ISD data into RDF, we reuse the data schema from our previous work~\cite{le2020pushing}, which employs the SSN/SOSA ontology~\cite{10.3233/SW-180320} to describe the metadata of sensors and the sensor readings in the ISD dataset.
%format of the sensor readings of the ISD dataset transformed by ~\cite{le2020pushing} using . 
%Mapping the values and attributes of each observation to the schema requires approximately 87 RDF triples. We selected the observation records of multiple weather stations, thereby representing different sizes of datasets. 
The process of mapping the values and attributes of each observation to the schema requires approximately 87 RDF triples. We have chosen observation records from multiple weather stations, thereby representing datasets of different sizes. The dataset is split and loaded into participant nodes with a reasonably balanced distribution of keys with the assumption that the P-Grid construction process has halted. Because the P-Grid exchange function was proven to balance the distribution of keys ~\cite{10.1007/3-540-44751-2_15}.
%Therefore, we assume that in the following experiments, the P-Grid construction process has halted, resulting in a reasonably balanced distribution of keys. 

%\begin{figure}
%    \centering
%    \includegraphics[width=\linewidth]{img/observation-schema.pdf}
%    \caption{The RDF schema of a sensor reading in the ISD dataset(from~\cite{le2020pushing})}
%    \label{fig:NOAA schema}
%\end{figure}

% -describe how you measure the metrics? what are they for? what you expect to see when you design such experiments? [Danh]

%% file: Chapter04/42Experiments.tex
\subsection{Experiments and Analysis}

\subsubsection{Exp1: QET of a Single Atomic Triple Pattern}
To initiate the study of our system's behaviour when responding to a SPARQL query, we measured the QET of a SPARQL query containing a single atomic triple pattern, as depicted in Listing~\ref{atomic}. Given that this query doesn't entail any join operations, this experiment aims to offer an analysis of how message passing within a P2P network influences the QET of such a P2P RDF engine.

\begin{lstlisting}[
caption={Atomic Triple Pattern - List all observations. },
label={atomic},
escapeinside=||
]
PREFIX sosa:<http://www.w3.org/ns/sosa/>
PREFIX rdf:<http://www.w3.org/1999/02/22-rdf-syntax-ns#>

SELECT ?observation
WHERE { ?observation rdf:type  sosa:Observation. %TP1} 
\end{lstlisting}

It is essential to note that the measurements we obtained here regarding the IO delay within our setup. Through a microbenchmark of network IO, we determined that the act of sending 1000 messages, each of 1KB in size, consumes approximately 1147 milliseconds. It's worth noting that the delay in local storage IO is notably minor in comparison, rendering it inconsequential when compared to the time taken for communication.

As mentioned in the previous section, the number of triples is divided approximately equally among the involved peers, indicating that the network achieved a balanced key distribution after multiple data exchange phases during P-Grid construction. With N participating nodes, the query initiator required at most log(N) hops to locate the final results.

\begin{figure}[ht!]
    \centering
    \begin{tikzpicture}
        \begin{axis}[
            xlabel={Size of ISD Dataset (the number of RDF triples)},
            ylabel={QET in milliseconds},
            grid style=dashed,
            ymajorgrids=true,
            width=\linewidth,
            height=5cm,
            legend pos=north west,
            legend style={font=\small},
            ytick={0, 1000,2000,3000,4000,5000,6000, 6500},
            xtick={1, 2, 3, 4, 5, 6, 7, 8},
            xticklabels={26K, 52K, 140K, 208K, 416K, 720K, 1M, 2M},
            enlarge x limits=0.1,
            enlarge y limits=0.1,
        ]
        \addplot[color=blue, mark=square] coordinates {(1, 935) (2, 1068) (3, 1437) (4, 1593) (5, 2290) (6, 2861) (7,3745) (8, 5651)};
        \addlegendentry{TP1, N=4}

        \addplot[color=red, mark=triangle] coordinates {(1, 900) (2, 1083) (3, 1523) (4, 1697) (5, 2341) (6, 3417) (7, 4136) (8, 6412)};
        \addlegendentry{TP1, N=8}

        \addplot[color=green, mark=diamond] coordinates {(1, 774) (2, 1097) (3, 1424) (4, 1757) (5, 2288) (6, 3017)(7, 3892) (8, 6308)};
        \addlegendentry{TP1, N=16}
        
        \end{axis}
    \end{tikzpicture}
    \caption{QET of Atomic Triple Pattern TP1 Using ISD Dataset. N is the number of peers in the system.}
    \label{fig:atomic}
\end{figure}
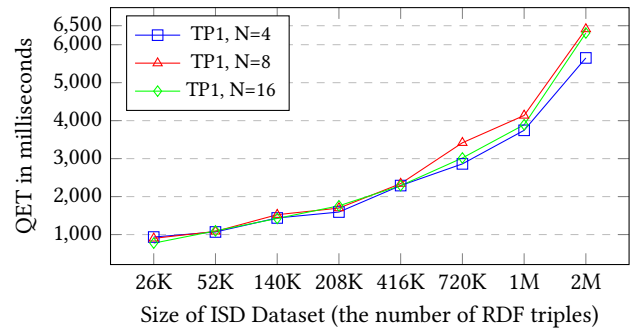

Under these data setup conditions, we varied the size of the ISD dataset to 26K, 52K, 140K, 208K, 416K, 720K, 1M, and 2M, as shown on the x-axis in Figure~\ref{fig:atomic}. Consequently, this led to varying numbers of RDF triples being returned for the atomic triple pattern: 2K, 4K, 10K, 16K, 31K, 54K, 75K, and 153K. It is worth noting that in this scenario, the size of the result set accounted for nearly 8\% of the total dataset. The number provided is significantly larger than the actual result size typically returned from a SPARQL query, which often falls below 1\% or even 0.1\%. We measured the QET by recording the time from the initiation of a request until the initiator received all matching results from the answering nodes. The test results for query execution time when responding to a single atomic triple pattern on different data scales in our setup are presented in Figure~\ref{fig:atomic}.

%Despite a retrieval rate of approximately 8\%,
As shown in Figure~\ref{fig:atomic}, our system experiences delays in searching and retrieving data, ranging from 1 to 6.5 seconds, across datasets comprising 26K to 2M triples. Throughout the querying process, the communication cost encompasses several factors, including the hops required to locate answering peers, the expense incurred as answering peers transmit messages containing possible block entries, the outlay for the query initiator to request matching RDF triples for each block entry received from answering peers, and the cost for answering peers to send messages containing matching RDF triples.

Furthermore, the results shown in Figure~\ref{fig:atomic} highlight that QET is significantly influenced by the number of matching RDF triples returned. Increasing the dataset size leads to a considerable delay increase. In this context, the difference in QET across various network sizes is not very significant. Increasing the number of involved nodes results in slight delays. This is primarily due to the fact that, when considering datasets of the same size, the number of matching RDF triples remains constant, with only one or two hops added during the searching phase.

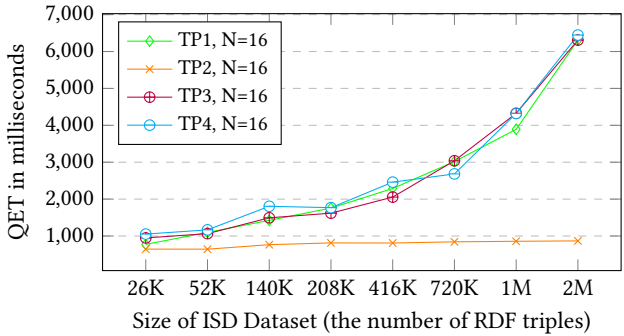
\begin{figure}[H]
    \centering
    \begin{tikzpicture}
        \begin{axis}[
            xlabel={Size of ISD Dataset (the number of RDF triples)},
            ylabel={QET in milliseconds},
            grid style=dashed,
            ymajorgrids=true,
            width=\linewidth,
            height=5cm,
            legend pos=north west,
            legend style={font=\small},
            ytick={0, 1000, 2000, 3000, 4000, 5000, 6000, 7000},
            xtick={1, 2, 3, 4, 5, 6, 7, 8},
            xticklabels={26K, 52K, 140K, 208K, 416K, 720K, 1M, 2M},
            enlarge x limits=0.1,
            enlarge y limits=0.1,
        ]

        \addplot[color=green, mark=diamond] coordinates {(1, 774) (2, 1097) (3, 1424) (4, 1757) (5, 2288) (6, 3017) (7, 3892) (8, 6308)};
        \addlegendentry{TP1, N=16}

        \addplot[color=orange, mark=x] coordinates {(1, 645) (2, 643) (3, 765) (4, 813) (5, 810) (6, 842) (7, 857) (8, 868)}; % sensor name unified:655010, triples around 4200
        \addlegendentry{TP2, N=16}

        \addplot[color=purple, mark=oplus] coordinates {(1, 948) (2, 1064) (3, 1493) (4, 1619) (5, 2055) (6, 3039)(7, 4325) (8, 6310)};
        \addlegendentry{TP3, N=16}

        \addplot[color=cyan, mark=halfcircle] coordinates {(1, 1052) (2, 1167) (3, 1805) (4, 1768) (5, 2460) (6, 2683)(7, 4322) (8, 6446)};
        \addlegendentry{TP4, N=16}
        \end{axis}
    \end{tikzpicture}
    \caption{QET of Atomic Triple Patterns TP1, TP2, TP3, TP4 Using ISD Dataset on 16 Pi4s. N is the number of peers.}
    \label{fig:atomic2}
\end{figure}

To gain further clarity on the impact of message passing quantity, we repeated the experiment using various triple patterns (TPs). To avoid redundancy, we're presenting results exclusively from our 16-node network. Figure~\ref{fig:atomic2} depicts the test outcomes utilizing a triple query pattern from the 2nd SPARQL query, employed in our second experiment (see Section~\ref{sec:exp2}). Given the similarity in the number of matched triples between TP3 and TP4 in the query shown in Listing~\ref{join}, and TP1 in the query presented in Listing~\ref{atomic}, the delays almost the same.

%Having established the paramount importance of the number of matching RDF triples, we strive to create a highly distributed environment to further investigate the QET of other types of atomic triple patterns from Listing~\ref{join}. 
%For example, ATP2 is of type(s, p, ?o), for which the number of matching RDF triples remains constant regardless of the size of various datasets. Thus, the retrieval rates for this query pattern vary, ranging from 0.2\% to 16\%. 
%result set size=4200, dataset size=[26K, 2M]
%These patterns will subsequently be employed in the evaluation of join query patterns, facilitating a comprehensive analysis of the system's performance in emulating distributed scenarios.

For TP2, we fixed the subject \%sensor\% to a specific sensor IRI, resulting in a fixed number of matched triples and returned results, even as the data scale increased. The QET remains consistent despite the growth in data size. Our system achieved the capability to return around four thousand results within less than a second in the context of a 16-node system.

\subsubsection{Exp2: QET of Complex Join Query Patterns} 
\label{sec:exp2}

%\hfill\\
A practical SPARQL query may contain multiple join operations, which will result in enormous execution time. Thus in this experiment, we further measure the QET of answering a query with multiple joins under uniform data distribution. Here we consider an example which contains a star-pattern join query as shown in Listing~\ref{join}. 

\begin{lstlisting}[
label={join},
caption={Join Query Pattern containing 3 atomic triple patterns -List the information of all observations made by a sensor.} ]
PREFIX sosa:<http://www.w3.org/ns/sosa/>
PREFIX rdf:<http://www.w3.org/1999/02/22-rdf-syntax-ns#>

SELECT ?obs ?featureOfInterest ?obsProperty
WHERE { 
%sensor%  sosa:madeObservation ?obs.               %TP2
?obs sosa:hasFeatureOfInterest ?featureOfInterest. %TP3
?obs sosa:observedProperty     ?obsProperty.       %TP4}
\end{lstlisting}

Using an ISD dataset of 26K triples, and a cluster of 16 Pi4s, the QET for the join query, as illustrated in Listing~\ref{join}, was found to be 11.15s. To extrapolate the execution time of join queries with uniform data distribution across various dataset sizes and network scales, we are prompted to employ synthetic data to execute an analogous join query.

To emulate a star pattern join query as in Listing~\ref{join}, the following join query is used :
%\begin{equation*}
%\begin{aligned}
%    q_1:(s_1, p_1, ?o_1) \textbf{JOIN} q_2:(?s_2, p_2, ?o_2) \textbf{ON} q_1.o_1=q_2.s_2 \\
%    \textbf{JOIN} q_3:(?s_3, p_3, ?o_3) \textbf{ON} q_1.o_1=q_3.s_3
%\end{aligned}
%\end{equation*}

\begin{lstlisting}[frame=none,escapeinside=||,basicstyle=\normalsize,label={abc}]
    |$q_1:(s_1, p_1, ?o_1)$ \textbf{JOIN} $q_2:(?s_2, p_2, ?o_2)$ \textbf{ON} $q_1.o_1=q_2.s_2$|
               |\textbf{JOIN} $q_3:(?s_3, p_3, ?o_3)$ \textbf{ON} $q_1.o_1=q_3.s_3$|
\end{lstlisting}

\begin{figure}[htbp]
    \centering
    \includegraphics[width=\linewidth]{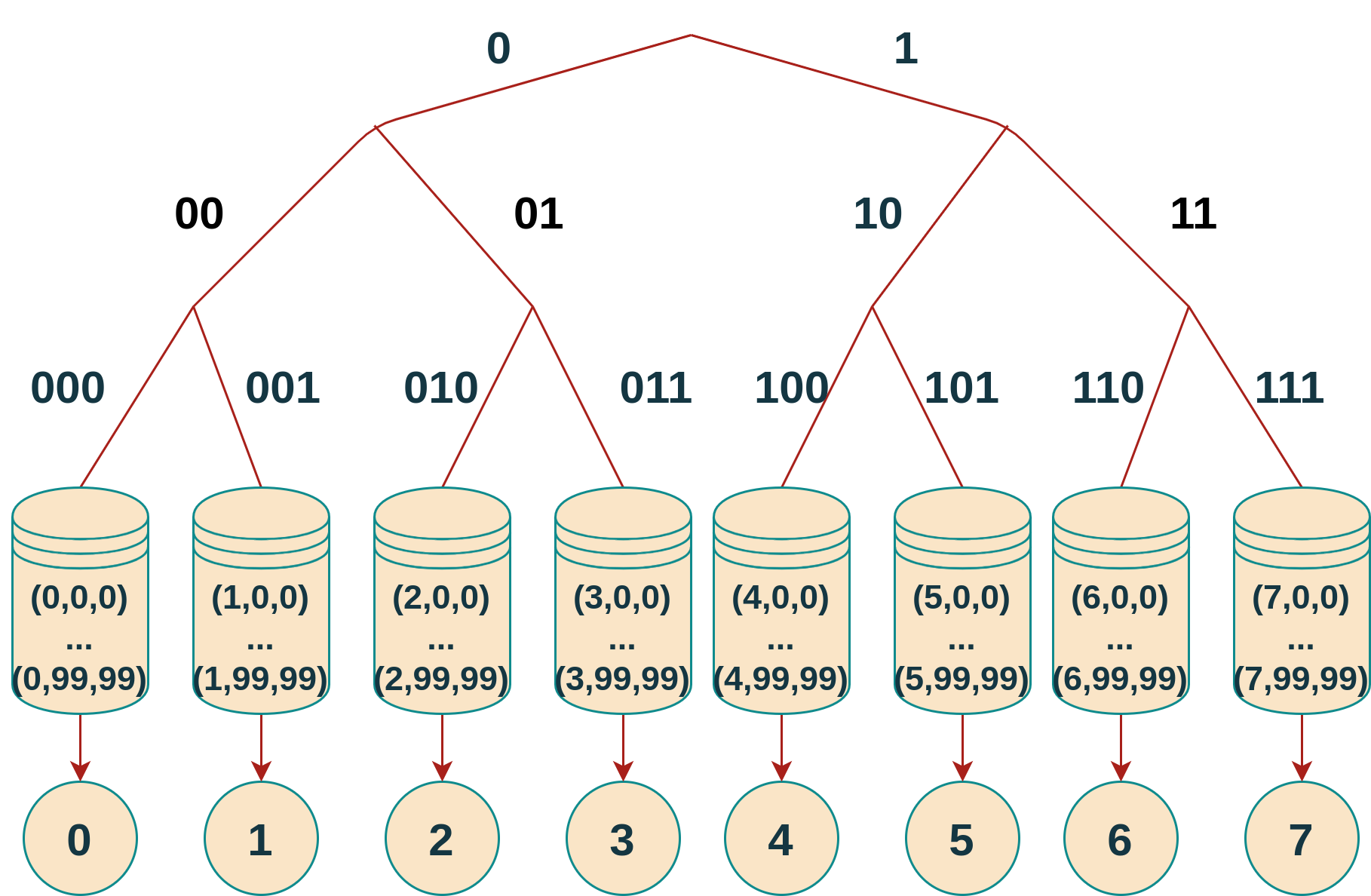}
    \caption{Example of two join operations with uniform data distribution. Each peer has $10^{4}$ tuples. Peer 0 initiates the join query.}
    \label{fig:2joins}
\end{figure}

\begin{figure}[htbp]
    \centering
    \includegraphics[width=0.8\linewidth]{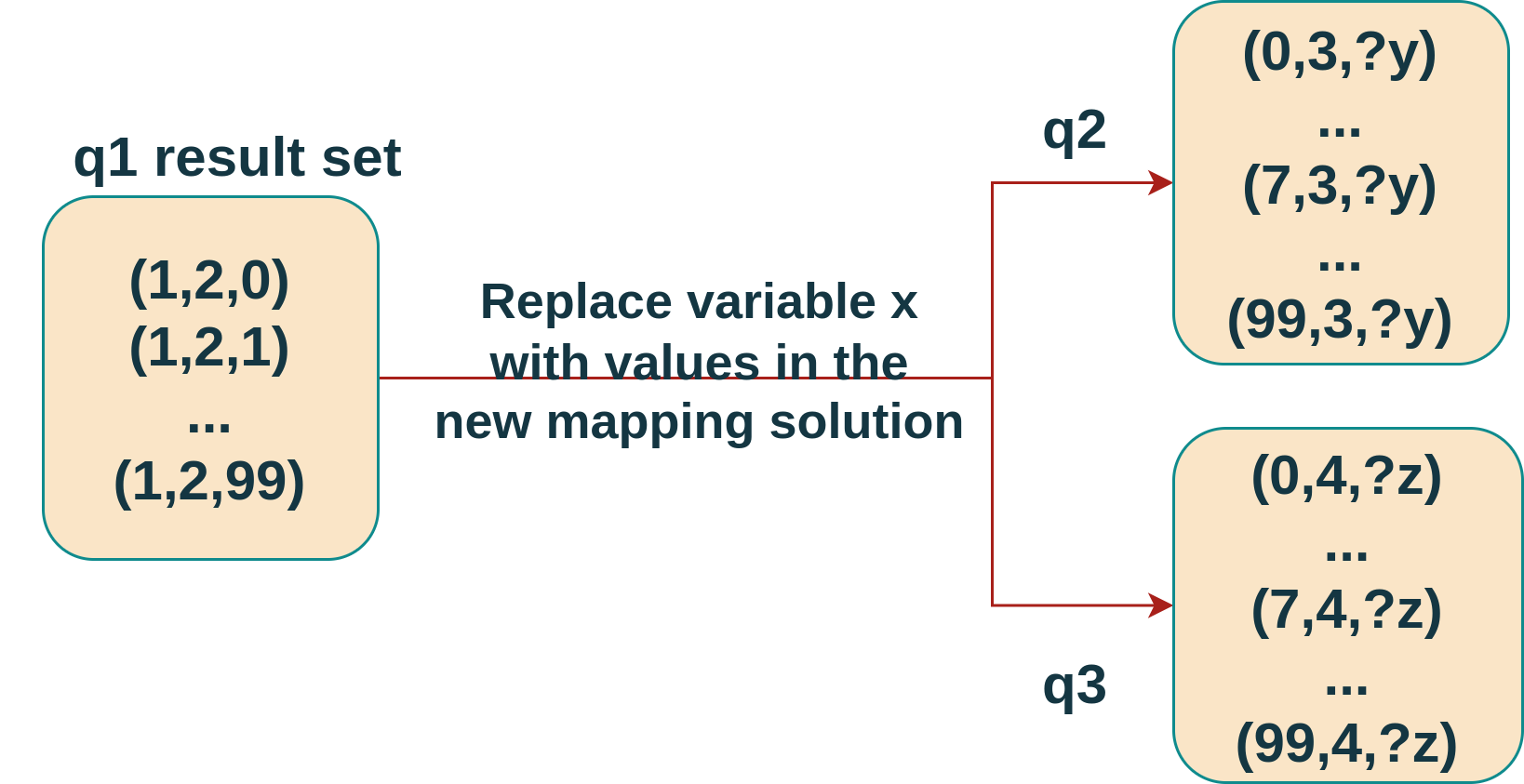}
    \caption{Process of bind join among $q_1$, $q_2$ and $q_3$. $q_2$ and $q_3$ share a common variable $x$ with $q_1$, making it possible for $q_2$ and $q_3$ to replace $x$ with real values in parallel. }
    \label{fig:join_process}
\end{figure}

In Figure~\ref{fig:2joins}, each peer stores an equal number of tuples, thereby demonstrating a well-balanced virtual P-Grid trie upon completion of construction. We consider the queries $q_1:(1, 2, ?x)$, $q_2:(?x, 3, ?y)$, and $q_3:(?x, 4, ?z)$ in Figure~\ref{fig:join_process}. As the join algorithm in~\cite{le2018rdf4led}, a mapping solution is kept and sent to each triple pattern of the graph pattern throughout the join process. We assume $q_1$ is initially visited, resulting in the variable $x$ and its corresponding values being added to the mapping. The new mapping will be sent to visit the other two triple patterns. Since $q_2$ and $q_3$ both contain variable $x$, replacing $x$ with each real value from the mapping solution and executing $q_2$ and $q_3$ in parallel becomes feasible, leading to reduced waiting time.  The retrieval rate for each answering node is fixed at 1 per cent of its local storage capacity.

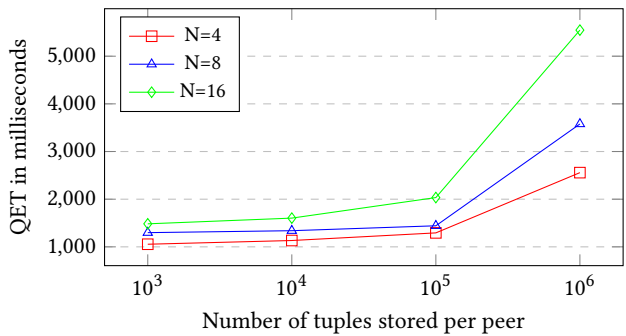
\begin{figure}[htbp]
    \centering
    \begin{tikzpicture}
    \begin{semilogxaxis}[
        width=\linewidth,
        height=5cm,    
        xlabel={Number of tuples stored per peer},
        ylabel={QET in milliseconds},
        legend pos= north west,
        legend style={font=\small},
        ytick={1000,2000,3000,4000, 5000, 6000},
        ymajorgrids=true,
        grid style=dashed,
        minor tick style={draw=none}]
    \addplot[
    color=red,
    mark=square,
    ]
    coordinates {
    (1e3, 1055)(1e4, 1133)(1e5, 1293)(1e6, 2557)
    };
    \addlegendentry{N=4}

    \addplot[
    color=blue,
    mark=triangle,
    ]
    coordinates {
    (1e3, 1298)(1e4, 1339)(1e5, 1442)(1e6, 3578)
    };
    \addlegendentry{N=8}

    \addplot[
    color=green,
    mark=diamond,
    ]
    coordinates {
    (1e3, 1483)(1e4, 1603)(1e5, 2034)(1e6, 5547)
    };
    \addlegendentry{N=16}
    \end{semilogxaxis}
    \end{tikzpicture}
    \caption{QET of Multiple Join Operations With Uniform Data Distribution. N is the number of peers in the system. }
    \label{fig:2joinsdata}
\end{figure}

Figure~\ref{fig:2joinsdata} presents the test results. As anticipated, the query execution time increases with the number of answering nodes and the storage size of each peer. 

The figure illustrates that there is a direct correlation between the execution time of the join query and the number of peers participating in the query. This suggests that the more peers are involved in the join query, the longer it takes to complete the query due to the increased communication overhead. Furthermore, a significant rise in execution time is observed when the number of tuples per peer reaches 1M. However, when the number of tuples per peer remains below $10^{5}$, the execution time shows little variation. This phenomenon may be attributed to the longer search time required for each answering peer in its local storage with a substantially larger dataset, resulting in an increased number of messages in transit.

%% file: 06Conclusion.tex
\section{Conclusion}
\label{seciton6}
The proposed approach has the potential to advance the field of RDF data management in IoT edge devices in terms of enabling effective integration of IoT data through semantic interoperability. We realised our approach as a distributed RDF engine by integrating two related works in the field: RDF4Led and P-Grid. 
Leveraging the advances of the two systems, our implementation preserves the two-layer storage structure from RDF4Led and the access structure of P-Grid to enable storing and querying RDF data on IoT devices with limited resources. 
We implemented the system using part of the source code from RDF4Led and P-Grid. Furthermore, we designed a set of experiments to evaluate the performance of the implementation in a P2P system using up to 16 Raspberry Pi 4 devices. The measurement and analysis of the time taken to search and join operations showed that our system is able to operate with different data sizes (up to 10 million per node).

The results presented in this paper pave the way for future research in semantic data processing on P2P networks at the edge of IoT. 
Our work contributes to the development of distributed RDF data stores and provides a foundation for future research on optimising query processing and exploring new data availability and replication techniques. 
The possible directions for extending this work can be to investigate new techniques of load imbalance caused by node departures or failures and data updates in a P2P system. Integrating our system with Saturn~\cite{saleem2019representative}, an overlay architecture on P-Grid, can enhance load distribution and fault tolerance. 
For multiple join queries, efficient management of intermediate results is crucial to mitigate network delays and I/O costs. Distributing join operators across nodes can optimise performance. This \textit{task assignment problem} has been addressed by certain papers~\cite{operator}~\cite{1617417} using a decentralised algorithm that progressively refines the placement of operators towards an optimal placement. 

% Another direction is to assess the performance of query processing of the proposed system under more realistic environments. For example, peers in real P2P networks may be unable to communicate directly due to a topology mismatch between the overlay and the underlying physical topology~\cite{topologyaware}~\cite{rostami2007topology}. This will cause a large volume of redundant traffic and delays. To address this, conducting more extensive experiments with diverse network topologies and various traffic intensities and incorporating topology awareness models into the overlay networks would be beneficial. Furthermore, the use of data compression techniques could expedite message passing speed. 

% Additionally, it would be worthwhile to consider the optimisation of join operations. We have only estimated the query execution time for a star-pattern join query in our experiments without consideration of a chain-pattern join query. If we adopt the join algorithm from RDF4Led, it is hard for a current node to know when the complete result set for a query pattern will be, which relies heavily on the number of answering nodes in the whole network. To address the trade-off between the completeness of the query result set and the execution time, setting a timeout for each query pattern with a prediction of execution time may be appropriate. 